# Review of the application piezoelectric actuators for SRF cavity tuners

Yuriy Pischalnikov, Crispin Contreras-Martinez

*Fermi National Accelerator Laboratory, Batavia, IL, USA*

pischaln@fnal.gov

**Abstract**

Modern particle accelerators and high-energy physics experiments that deployed up to several hundred of accelerating superconducting RF cavities require accurate frequency control. This is achieved by using cavity tuners typically actuated with the piezoelectric ceramic actuators. Piezoelectric ceramic actuators have become "standard" components of the SRF cavity tuner and depending on the application could be operated in different environments: in air, at cryogenic temperature, in vacuum, and submerged in liquid helium. Different applications place different requirements on the piezo actuators, but the important parameters common to all applications are the lifetime and reliability of the actuators. Several programs targeting the development of reliable piezo actuators are presented in this contribution.

## 1. Introduction

Modern particles accelerators such as Eu-XFEL/DESY [1], LCSL-II/SLAC [2], and ESS [3] are large in size. They are built for fundamental physics research and deploy hundreds of accelerating elements known assuperconducting radio frequency (SRF) cavities made from niobium. During operation in an accelerator, the SRF cavity frequency must be actively adjusted by changing length of the ~1 m long cavity on the level nanometers. Frequency tuning systems (or simply tuners) are designed and operated to perform this function (Figure 1) [4].

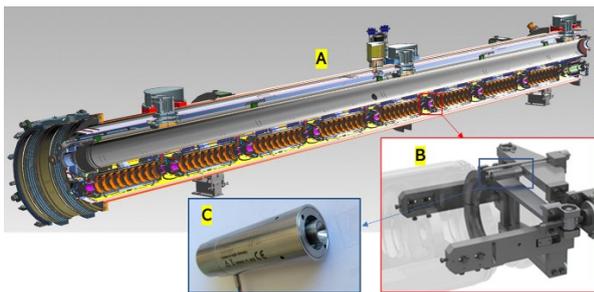

**Figure 1:** 3D picture of the LCLS-II accelerator cryomodule with 8 elliptical SRF cavities (A). Each cavity has tuner system that adjust frequency of SRF cavity to resonance (B). To compress cavity (to change cavity frequency) encapsulated piezo actuator P-844K075 has been deployed (C).

The SRF cavity tuner system is a combination of a tuning mechanism and an actuator. Typically, such systems are a mechanical frame with piezoelectrical elements serving as actuators to compress or stretch cavity. The ability of the piezo actuator to generate large forces (approx. 4 kN for stack with cross-section 10*10 mm$^2$) and capability to deliver stroke up 10's of um (with almost unlimited resolution) and withstand pressure up to 200 MPa make these actuators as good choice to be deployed in SRF cavity tuners.

Recently thousands of piezo actuators deployed as fast/fine tuning element in several large SRF linac that in operation or close to be in operation. In large machine (Eu-XFEL, LCLS-II, ESS) piezo actuators typically deployed close to SRF cavity, inside SRF cryomodule at insulated vacuum volume and at cryogenic working temperature. Reliability of the piezo actuators became most critical parameter considering complexity and cost of replacement in case of failure.

## 2. Measurements of piezo stroke versus temperature

As part of the LCLS-II cryomodules commissioning, after cooling cavities to T=2 K, the detuning of each cavity with piezo actuators has been measured [4]. The piezo actuators stroke when cooling to a range of temperature between T=4 K to T=10 K is around 10 um or almost 2-3 times larger than expected from data presented at several papers [5,6] and companies' catalogues [7, 8]. To address this discrepancy the stroke of actuators P-844K075 [9], made from PZT (Lead Zirconate Titanate) 10*10*36 mm$^3$ PICMA® [10], has been measured at different temperature. The DarkPhoton Search experiment setup [11] has been used to conduct measurements (Figure 2). Stroke of the actuator P-844K075, measured before installation into setup at T=295K and V$_{piezo}$=120 V, was 36 um.

The piezo actuator, when V$_{piezo}$=120 V is applied, retuned the single cell 1.3GHz SRF cavity by compressing the cavity. Cavity retuning (frequency shift) was measured with a NWA (Network Analyzer) (Figure 3). Cavity was installed into special facility and cooled/submerged with liquid Helium to T=4K. Cavity retuning (or the piezo stroke) has been measured at several points (at different temperature) after liquid helium was evaporated and cavity/tuner/piezo system slowly warmed up to room temperature (during approximately 24 hours). The temperature of the piezo ceramic was measured with an RTD (CERNOX) installed on the surface of the piezo actuator. The SRF cavity parameter df/dL=2.3 kHz/um (the cavity's frequency change (df) versus cavity length change (dL)) is not dependent from cavity temperature and thus allowed the cavity to be used as high-resolution sensor for piezo stroke at different temperatures. At a temperature range from T=2 K to T=9 K, when Nb cavity is in the superconducting phase, quality factor of the SRF cavity is high and bandwidth is narrow. The cavity's narrow bandwidth led to cavity's retuning accuracy approximately several tens of Hz. The accuracy of piezo stroke measurements through of the cold cavity detuning is in nanometers range. And when cavity go to "normal conducting stage" quality factor will drop and accuracy of the piezo stroke measurements will be 1 micrometer or better.

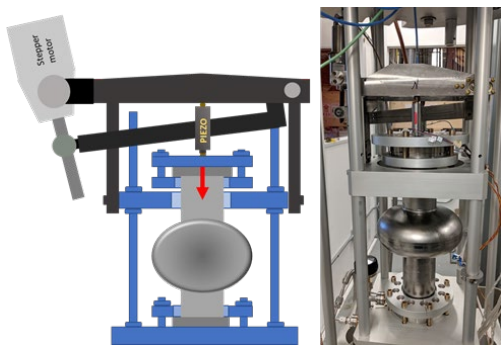

**Figure 2:** DarkPhoton search experiment setup, with single cell 1.3 GHz SRF cavity and piezo tuner equipped with piezo actuator, used to measure piezo response versus temperture of piezo ceramic.

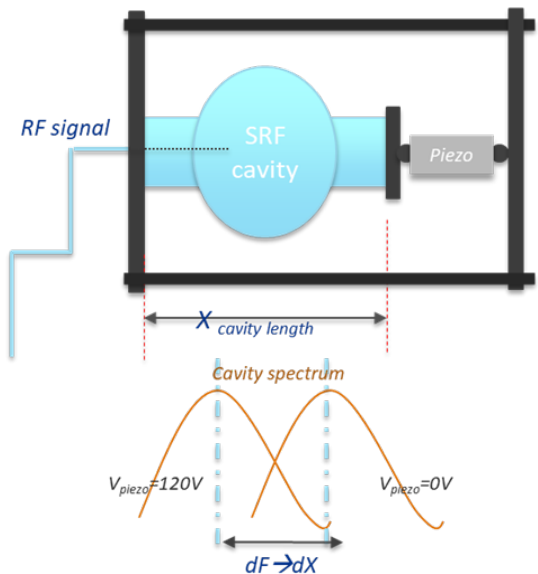

**Figure 3:** Schematics of setup used for measurements of piezo stroke through retuning/compression of the SRF cavity. Changes in the cavity frequency when the piezo was operated with $V_{piezo}$=0V and $V_{piezo}$=120V are then converted to piezo stroke.

But main objectives were not direct measurements of piezo stroke in micrometers rather find dependence of the piezo stroke from room temperature to cryogenic temperature. The main contributions in the measurements errors were due to changes in the cavity and tuner stiffness vs temperature and accuracy in the frequency detuning measurements when cavity was at higher temperature. The estimated measurements errors are below 10%. The results are shown in Figure 4. The piezo response (cavity detuning range) when SRF cavity cools down to T=4 K, was just 3-4 times lower than at "room temperature" rather 10-13 times as reported by other authors [5,6].

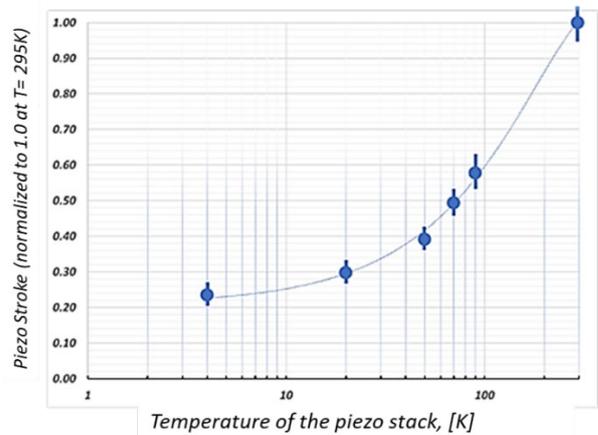

**Figure 4:** Dependency of PICMA ® piezo actuator stroke versus temperature of the piezoceramic stack. Piezo actuator stroke at room temperature (T=295K) normalized to 1.0.

## 3. Challenges for operation of piezo actuators at cryogenic temperature and vacuum

When the SRF cavity is operated in RF-pulse mode it is subject to dynamic Lorentz force detuning (LFD) up to several kHz. To compensate this level of cavity detuning, the piezo needs to be operated with stimulus pulse with amplitude up to 120-150 V. Power dissipation inside the piezo ceramic could be reach a level of 0.1 W up to 1 W. As demonstrated by several teams [13, 14], the temperature in the centre of the piezo stack when operated inside vacuum at high dynamic rate could be quickly raised by ΔT~200 K that could lead to piezo failure. Heating of the piezo when operating at the high dynamic rate that is required for LFD compensation could therefore significantly decrease the lifetime of the actuator [13,14,15]. The FNAL and Physik Intrumente (PI) teams conducted an R&D program to develop a novel piezo actuator for operation at cryogenic temperature, inside vacuum and at high dynamic rate. The goal was to develop an encapsulated piezo actuator that removes heat from the surface of the piezo-ceramic stack and prevents positive feedback heating. The result of this R&D program is an actuator P-844K093 (Figure 5 and 6). The encapsulated and preloaded actuator utilized the PICMA® stack of size 10*10*36mm$^3$. Copper foam was selected to remove heat from the side surfaces of the PICMA® stack. Copper foam installed between piezo stack and plate made form aluminium nitride. Aluminium nitride is a good dielectric with excellent heat transfer properties. This material isolated the outside encapsulation from high voltage while at the same time allowing for efficient heat extraction. For efficient operation a heat sink must be connected to actuator copper plate (attached to aluminium nitrade plate) and anchored to a cryomodule cryogenic pipe ( at T=2 K, T=4 K or T=77 K).

Dielectric heating measurements of the PICMA stacks in the two actuators P-844K075 and P-844K093, when operated inside vacuum and at cryogenic temperature were performed at designated facility (Figure 7) [14, 15]. Heating of both actuators when operated with the same sinewave stimulus pulse (*f*=100 Hz, $V_{piezo}$=100 V) are presented on Figure 8. Temperature of the actuator P-844K075 increased from T=20 K up to T~110 K during 1.5 hours, when stimulus pulse applied to the piezo actuator. At the same time temperature of the P-844K093 (piezo with copper foam) warmed up from 10 K by ∆T=7 K just during first 10 minutes of operation and not increased beyond 17 K. Piezo actuator P-844K093 (with copper foam) demonstrated the significant heat removal capability, generated by piezo stacks. The heat sink was attached to piezo actuator and T=4 K plate. Efficient transfer of the heat from piezo will prevent overheating and will significantly increase lifetime of piezo actuator.

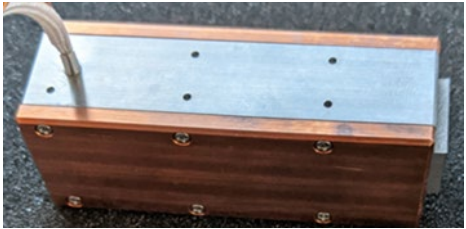

**Figure 5:** Picture of the novel piezo actuator P-844K093.

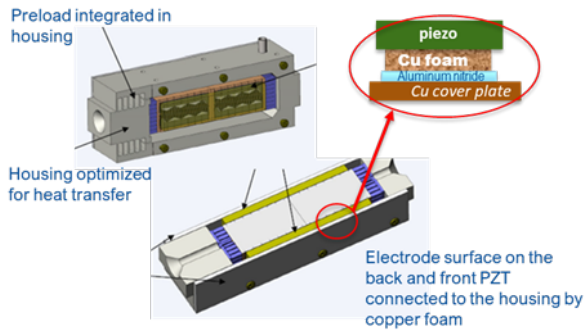

**Figure 6:** The details of design for actuator P-844K093, developed to operate at high amplitude and high dynamic rate at cryogenic temperature and inside vacuum.

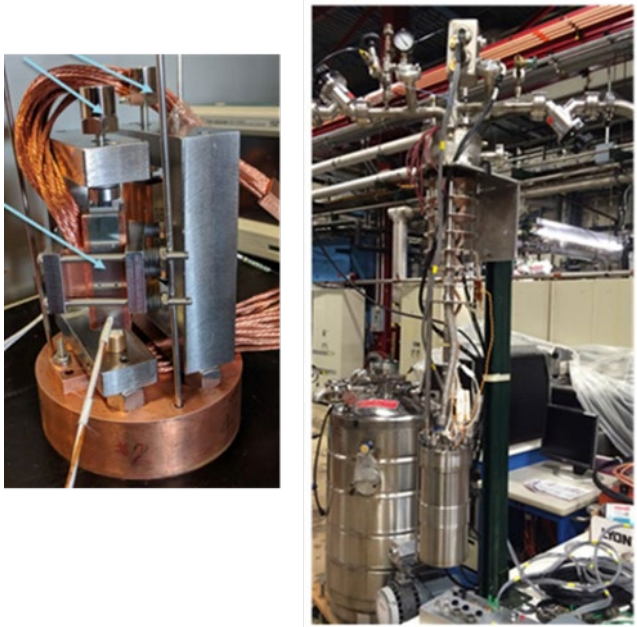

**Figure 7:** Facility for the piezo actuators Accelerated Lifetime Testing. Right: Helium Dewar and vacuum insert for piezo. Left: The novel piezo actuator P-844K093 with heat sink attached mounted on the heavy copper plate to be installed into vacuum insert.

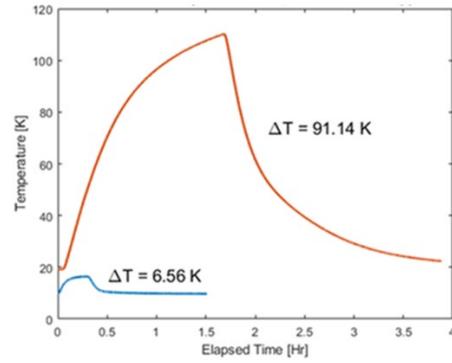

**Figure 8:** Results of the piezoceramic stacks (P-844K075(red) and P-844K093 (blue)) temperture increase after heating up with sinewave stimulus pulse ($V_{pp}$=100 V and f=100 Hz).

## 4. Summary

Results of the tests presented in this paper demonstrated that the stroke of actuator made from PZT piezo ceramics, when cooled to T~4K will decrease to ~1/4 of the stroke at room temperature. In a previous studies and inside companies' catalogues reported drop of the PZT stroke when cooldown to T=4K in ~10 times. These results will lead to more efficient design of piezo tuners for future very large size SRF accelerators (like ILC [ 16]), where tens of thousands of piezo actuators will be deployed.

A novel piezo actuator P-844K093 has been developed to increase the realiabity of piezo actuators in the SRF cavity tuners which operate in RF pulse mode. Developed design allows for efficient removal of heat from the piezo ceramic and preventation of the positive feedback heating. The tests were conducted by running piezo actuators at high voltage amplitude and high dynamic rate inside vacuum and at temperature range of T=10 K to T=20 K. The temperature increase of P-844K093 due to piezoelectric stack dielectric heating was reduce by a factor of 15 compared to standard (P-844K075) PZT actuator.

Another approach to mitigate dielectric heating of piezo actuators is to develop actuators from different piezoelectric ceramic materials, which will generate less dialectric heating. There are requests for development of newest low heat generating piezoceramic from scientists working in the field of Dark Matter search experiments. The Dark Matter experiments deployed inside Dilution Refrigerators at temperature range T~10-20 mK with heat removal capacities in the level of microwatts.